\documentclass[smallextended,numbook]{svjour3}
\usepackage{graphicx}

\newcommand{\beq}{\begin{equation}}
\newcommand{\beqa}{\begin{eqnarray}}
\newcommand{\eeq}{\end{equation}}
\newcommand{\eeqa}{\end{eqnarray}}
\newcommand{\abs}[1]{\left\vert#1\right\vert}
\newcommand{\bigmean}[1]{\left\langle#1\right\rangle}
\newcommand{\col}{{\mathrm{s}}}
\renewcommand{\d}{{\rm d}}
\newcommand{\de}{{(2)}}
\newcommand{\ds}{\displaystyle}
\newcommand{\e}{{\rm e}}
\newcommand{\eps}{\varepsilon}
\newcommand{\erfc}{\mathop{\rm erfc}\nolimits}
\newcommand{\euler}{{\gamma_{\scriptscriptstyle{\rm E}}}}
\newcommand{\frad}[2]{\ds{\frac{#1}{#2}}}
\newcommand{\frat}[2]{\ts{\frac{#1}{#2}}}
\renewcommand{\i}{{\rm i}}
\newcommand{\mean}[1]{\langle#1\rangle}
\newcommand{\prob}{\mathop{\rm Prob}\nolimits}
\newcommand{\sca}{{\mathrm{scal}}}
\newcommand{\sg}{{\mathrm{sing}}}
\newcommand{\st}{{\mathrm{stat}}}
\newcommand{\ts}{\textstyle}
\newcommand{\var}{\mathop{\rm var}\nolimits}
\newcommand{\w}{\widehat}
\newcommand{\A}{{(\mathrm{A})}}
\newcommand{\B}{{(\mathrm{B})}}
\newcommand{\C}{{\mathrm{C}}}
\renewcommand{\Re}{\mathop{\rm Re}\nolimits}

\journalname{Journal of Statistical Physics}

\begin{document}

\title{Finite-time fluctuations in the degree statistics of~growing networks}

\author{C.~Godr\`eche \and H.~Grandclaude \and J.M.~Luck}

\institute{Institut de Physique Th\'eorique, IPhT, CEA Saclay,
and URA 2306, CNRS, 91191 Gif-sur-Yvette cedex, France\\
\email{claude.godreche@cea.fr, helene.grandclaude@cea.fr,
jean-marc.luck@cea.fr}}

\date{}

\maketitle

\begin{abstract}
This paper presents a comprehensive analysis of the degree statistics
in models for growing networks where new nodes enter one at a time
and attach to one earlier node according to a stochastic rule.
The models with uniform attachment,
linear attachment (the Barab\'asi-Albert model),
and generalized preferential attachment with initial attractiveness are
successively considered.
The main emphasis is on finite-size (i.e., finite-time) effects,
which are shown to exhibit different behaviors
in three regimes of the size-degree plane:
stationary, finite-size scaling, large deviations.
\end{abstract}

\section{Introduction}
\label{intro}

Complex networks have attracted much attention over the last decades.
They provide a natural setting to describe many phenomena
in nature and society~\cite{abrmp,doro1,doro2,blmch,cup}.
One of the salient features of most networks,
either natural and artificial, is their scalefreeness.
This term refers to the broad degree distribution exhibited by these networks.
The probability that a node
has degree $k$ (i.e., is connected to exactly $k$ other nodes)
is commonly observed to fall off as a power law:
\beq
f_k\sim k^{-\gamma}.
\label{f}
\eeq
This power-law behavior,
which holds in the limit of an infinitely large network,
will be referred to hereafter as `stationary'.
The exponent usually obeys $\gamma>2$,
so that the mean degree of the infinite network is finite.
Growing networks with a preferential attachment rule,
such as the well-known Barab\'asi-Albert (BA) model~\cite{ba1,ba2},
have received a considerable interest,
as they provide a natural explanation for the observed scalefreeness.
The observation that preferential attachment generates
a power-law degree distribution actually dates back
to much earlier works~\cite{simon,price}.

Scalefree networks, being chiefly characterized by the exponent $\gamma$
of their degree distribution,
are therefore somewhat similar to equilibrium systems
at their critical point.
As a consequence,
finite-size (i.e., finite-time) effects can be expected to yield
important corrections to the asymptotic or stationary form~(\ref{f})
of the degree distribution.
These effects are one of the possible causes of the cutoff phenomenon
which is often observed in the degree distribution of real networks~\cite{bpv}.
More precisely,
the largest degree $k_\star(n)$ of a scalefree network at time $n$
can be estimated by means of the following argument of extreme value statistics:
it is such that the stationary probability
of having $k\ge k_\star(n)$ is of order $1/n$.
The largest degree thus grows as a power law~\cite{bpv,dms1}:
\beq
k_\star(n)\sim n^\nu,\quad\nu=\frac{1}{\gamma-1}.
\label{kstar}
\eeq
This growth law is always subextensive,
because one has $\gamma>2$, so that $\nu<1$.
The cases $2<\gamma<3$ (i.e., $1/2<\nu<1$)
and $\gamma>3$ (i.e., $0<\nu<1/2$)
however correspond to qualitative differences,
especially in the topology
and in the various dimensions of the networks~\cite{bck}.

The goal of this article is to provide a systematic analysis
of the degree statistics
of growing network models at a large but finite time $n$.
Both the age-resolved distribution $f_k(n,i)$
of the degree of node $i$ at a later time $n$
and the distribution $f_k(n)$ of an unspecified node at time $n$
will be considered throughout.
Several works have already been devoted to this problem,
both for growing networks
with preferential attachment~\cite{dms1,dms2,krl,kr1,kr2,ws,mj}
and for related models of random graphs and other structures~\cite{kk,cb}.
The present work aims at being systematic in the following three respects:

\noindent $\bullet$ {\it Models.}
This work is focussed onto growing network models where
a new node enters at each time step,
so that nodes can be labeled by their birth date~$n$,
i.e., the time they enter the network.
Node $n$ attaches to a single earlier node ($i=1,\dots,n-1$)
with probability $p_{n,i}$.
The attachment probabilities and the initial configuration
entirely define the model.
The network thus obtained has the topology of a tree.
The degrees $k_i(n)$ of the nodes at time $n$ obey the sum rule
\beq
\sum_{i=1}^n k_i(n)=2L(n),
\label{sumr}
\eeq
where $L(n)$ is the number of links of the network at time $n$.

We will successively consider the following models:

\noindent -- {\it Uniform attachment} (UA) (Section~2).
The attachment probability is independent of the node,
i.e., uniform over the network.
This model is not scalefree.
Its analysis serves as a warming up for that of the subsequent models.

\noindent -- {\it Barab\'asi-Albert} (BA) {\it model} (Section~3).
The attachment probability is proportional
to the degree $k_i(n)$ of the earlier node.
This well-known model~\cite{ba1,ba2} is scalefree,
with exponents $\gamma=3$ and $\nu=1/2$.

\noindent -- {\it General preferential attachment} (GPA) (Section~4).
The attachment probability is proportional
to the sum $k_i(n)+c$ of the degree of the earlier node
and of an additive constant $c>-1$.
This parameter,
representing the initial attractiveness of a node~\cite{dms1},
is relevant as it yields the continuously varying exponents
$\gamma=c+3$ and $\nu=1/(c+2)$.
The BA and UA model are respectively recovered when $c=0$ and $c\to\infty$.

\noindent $\bullet$ {\it Regimes.}
For each model, the following three regimes will be considered:

\noindent -- {\it Stationary regime} ($k\ll k_\star(n)$).
The degree distribution is essentially given by its stationary form~(\ref{f}),
to be henceforth denoted by $f_{k,\st}$,
in order to emphasize its belonging to the stationary regime.

\noindent -- {\it Finite-size scaling regime} ($k\sim k_\star(n)$).
In the scalefree cases, the degree distribution obeys
a multiplicative finite-size scaling law of the form
\beq
f_k(n)\approx f_{k,\st}\,\Phi\!\left(\frac{k}{k_\star(n)}\right).
\label{fssdef}
\eeq

\noindent -- {\it Large-deviation regime} ($k_\star(n)\ll k\sim n$).
The degree distribution is usually exponentially small in $n$.

\begin{table}
\caption{Various characteristics of the network for both initial conditions.
The listed results hold irrespective of the attachment rule.}
\label{tabledef}
\begin{tabular}{l|l|l}
\hline\noalign{\smallskip}
Initial condition & Case~A & Case~B\\
\noalign{\smallskip}\hline\noalign{\smallskip}
Topology & tree & rooted tree\\
Number of links at time $n$ & $L^\A(n)=n-1$ & $L^\B(n)=n-1/2$\\
Mean degree at time $n$ & $\mean{k^\A(n)}=2-2/n$ & $\mean{k^\B(n)}=2-1/n$\\
\noalign{\smallskip}\hline\noalign{\smallskip}
Degrees at time $1$ & $k^\A_1(1)=0$ & $k^\B_1(1)=1$\\
and generating polynomials& $F^\A_1(x)=1$ & $F^\B_1(x)=x$\\
\noalign{\smallskip}\hline\noalign{\smallskip}
Degrees at time $2$ & $k^\A_1(2)=k^\A_2(2)=1$ & $k^\B_1(2)=2$, $k^\B_2(2)=1$\\
and generating polynomials& $F^\A_2(x)=x$ & $F^\B_2(x)={\textstyle\frac{1}{2}}x(x+1)$\\
\noalign{\smallskip}\hline
\end{tabular}
\end{table}

\noindent $\bullet$ {\it Initial conditions.}
We will consider the following two initial conditions:

\noindent -- {\it Case~A.}
The first node appears at time $n=1$ with degree $k_1(1)=0$.
This prescription is natural because the first node initially has no connection.
All subsequent nodes appear with degree $k_n(n)=1$.
In particular, at time $n=2$ the second node connects to the first one,
so that $k_1(2)=k_2(2)=1$.
The configuration thus obtained is the dimer configuration
used e.g.~in~\cite{kr1,kr2}.
At time $n$, the network has $L(n)=n-1$ links.
It has the topology of a tree.

\noindent -- {\it Case~B.}
The first node now appears at time $n=1$ with degree $k_1(1)=1$.
This formally amounts to saying that this node is connected to a root,
which does not belong to the network.
It is natural to associate half a link to this fictitious connection.
At time $n=2$ the second node connects to the first one,
so that $k_1(2)=2$ and $k_2(2)=1$.
At time $n$, the network has $L(n)=n-1/2$ links.
It has the topology of a rooted tree.

\begin{figure}
\begin{center}
\includegraphics[width=.8\linewidth]{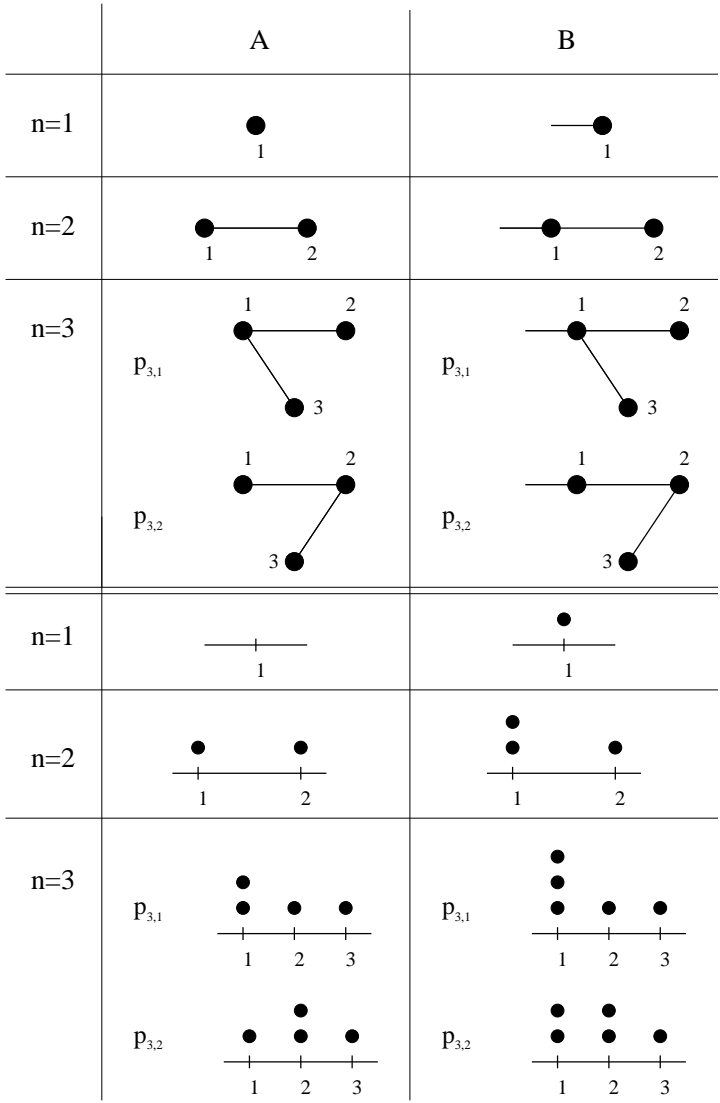}
\caption{\label{figdef}
First three steps of the construction of the network (upper panel)
and corresponding interacting particle representation (lower panel)
for both initial conditions.}
\end{center}
\end{figure}

Table~\ref{tabledef} summarizes various characteristics of the network
for both initial conditions,
whereas Figure~\ref{figdef} illustrates the first three steps of the network
construction.
The upper panel shows the networks with their nodes and links.
The lower panel shows the corresponding representation
as an interacting particle system,
where each node is viewed as a site occupied by
a number of particles equal to its degree.
The total number of particles in the system is therefore $2L(n)$.
The information about the topology of the network,
and especially about the genealogy of the nodes,
is lost in the interacting particle representation,
but this information will not be used in the present study
which is focussed on the statistics of degrees.

\section{The uniform attachment (UA) model}
\label{UA}

The uniform attachment (UA) model is the simplest of all:
the attachment probability is chosen to be uniform over all existing nodes.
This section is devoted to an analytical study of the distribution
of the degree of a fixed node and of an unspecified node,
exactly taking into account fluctuations, finite-time effects,
and the influence of the initial condition.

\subsection{Degree statistics of a fixed node}

We start with the study of the distribution
of the degree $k_i(n)$ of node $i$ at time $n$.
The node appearing at time $n\ge2$ links to any of the $n-1$ earlier nodes
($i=1,\dots,n-1$) with uniform probability
\beq
p_{n,i}=\frac{1}{n-1}.
\eeq

If we define the degree increment of node $i$ at a later time $j>i$ as
\beq
I_i(j)=k_i(j)-k_i(j-1)=\left\{\begin{array}{l l}1 &\mathrm{with\ probability\
}p_{j,i},\\0 &\mathrm{else},\end{array}\right.
\label{idef}
\eeq
the degree $k_i(n)$ of node $i$ at a later time $n$ is given by
\beq
k_i(n)=k_i(i)+\sum_{j=i+1}^n I_i(j),
\label{kinsum}
\eeq
with $k_i(i)=1$, except for $i=1$ in Case~A, where $k_1(1)=0$
(see Table~\ref{tabledef}).

The mean degree $\mean{k_i(n)}$ therefore reads ($i\ge2$)
\beq
\mean{k_i(n)}=1+\sum_{j=i+1}^n\frac{1}{j-1}
=H_{n-1}-H_{i-1}+1\approx\ln\frac{n}{i}+1,
\label{kave}
\eeq
where the harmonic numbers $H_n$ are defined in~(\ref{hardef}).

The distribution $f_k(n,i)=\prob\{k_i(n)=k\}$
can be encoded in the generating polynomial
\beq
F_{n,i}(x)=\bigmean{x^{k_i(n)}}=\sum_{k=1}^{n}f_k(n,i)x^k.
\eeq
As a consequence of~(\ref{kinsum}), we have
\beq
F_{n,i}(x)=x^{k_i(i)}\prod_{j=i+1}^n\bigmean{x^{I_i(j)}},
\label{fniprod}
\eeq
where the characteristic function of the degree increment $I_i(j)$
assumes the simple form
\beq
\bigmean{x^{I_i(j)}}=1+(x-1)p_{j,i}=\frac{x+j-2}{j-1},
\eeq
irrespective of $i$.
We thus get ($i\ge2$)
\beq
F_{n,i}(x)=\frac{x(i-1)!\Gamma(x+n-1)}{(n-1)!\Gamma(x+i-1)},
\label{fnires}
\eeq
whereas only $F_{n,1}(x)$ depends on the initial condition according to
\beq
F_{n,1}^\A(x)=\frac{\Gamma(x+n-1)}{(n-1)!\Gamma(x)},\quad
F_{n,1}^\B(x)=\frac{x\Gamma(x+n-1)}{(n-1)!\Gamma(x)}.
\eeq
Throughout the following, the superscripts $\A$ and $\B$ mark a result
which holds for a prescribed initial condition (Case~A or Case~B).

The product form~(\ref{fniprod}) implies that the generating polynomials
of node $i$ at times $n$ and $n+1$ obey the recursion
\beq
F_{n+1,i}(x)=\bigmean{x^{I_i(n+1)}}F_{n,i}(x)=\frac{x+n-1}{n}\,F_{n,i}(x).
\label{fniprec}
\eeq
The probabilities $f_k(n,i)$ therefore obey the recursion
\beq
f_k(n+1,i)=\frac{1}{n}\,f_{k-1}(n,i)+\left(1-\frac{1}{n}\right)f_k(n,i),
\label{fkdif}
\eeq
with initial conditions given in Table~\ref{tabledef}, i.e.,
\beq
f_k(i,i)=\delta_{k,1}\quad(i\ge2),\quad
f_k^\A(1,1)=\delta_{k,0},\quad f_k^\B(1,1)=\delta_{k,1}.
\label{f2init}
\eeq
The master equations~(\ref{fkdif}) can be directly written down
by means of a simple reasoning.
They provide an alternative way of describing
the evolution of the degree distribution of individual nodes.

The degree distribution encoded in~(\ref{fnires})
has the following characteristics.
The degree of node $i$ at time $n$ ranges from the
minimal value 1 to the maximal value $n+1-i$.
These extremal values occur with probabilities
\beq
f_1(n,i)=\frac{i-1}{n-1},\quad f_{n+1-i}(n,i)=\frac{(i-1)!}{(n-1)!}.
\eeq
The mean and the variance of the degree can be obtained
by expanding the result~(\ref{fnires}) around $x=1$, using
\beq
\mean{x^K}=1+(x-1)\mean{K}+\frac{1}{2}(x-1)^2
{\hskip -9pt}\underbrace{\mean{K^2-K}}_{\var{K}+\mean{K}^2-\mean{K}}+\cdots,
\label{devt}
\eeq
where $K$ is any random variable taking positive integer values.
We thus get
\beq
\matrix{
\mean{k_i(n)}=H_{n-1}-H_{i-1}+1,\hfill\cr\cr
\var{k_i(n)}=H_{n-1}-H^\de_{n-1}-H_{i-1}+H^\de_{i-1},\hfill
}
\label{kmom}
\eeq
where the harmonic numbers $H_n$ and $H^\de_n$ are defined in~(\ref{hardef}).
The above results hold irrespective of the initial condition.
The first one coincides with~(\ref{kave}).

In the scaling regime where both times $i$ and $n$ are large and comparable,
introducing the time ratio
\beq
z=\frac{n}{i}\ge1,
\label{zdef}
\eeq
the expressions~(\ref{kmom}) yield
\beq
\mean{k_i(n)}\approx\ln z+1,\quad\var{k_i(n)}\approx\ln z.
\label{kinsca}
\eeq

In deriving the above results,
we have used the asymptotic behavior
of the digamma function $\Psi(x)=\Gamma'(x)/\Gamma(x)$
and of the trigamma function~$\Psi'(x)$ as $x\to\infty$:
\beq
\Psi(x)=\ln x-\frac{1}{2x}+\cdots,\quad
\Psi'(x)=\frac{1}{x}+\frac{1}{2x^2}+\cdots,
\eeq
as well as their values at integers:
\beq
\Psi(n)=H_{n-1}-\euler,\quad
\Psi'(n)=\frac{\pi^2}{6}-H^\de_{n-1},
\eeq
where
\beq
H_n=\sum_{i=1}^n\frac{1}{i},\quad
H^\de_n=\sum_{i=1}^n\frac{1}{i^2}
\label{hardef}
\eeq
are the harmonic numbers of the first and second kind,
and $\euler$ is Euler's constant.

The entire degree distribution can be characterized in the scaling regime.
Equation~(\ref{fnires}) indeed yields
\beq
F_{n,i}(x)\approx x\,\e^{(x-1)\ln z},
\eeq
irrespective of the initial condition.
We recognize the generating function of a Poissonian distribution
with parameter $\lambda=\ln z$, up to a shift by one unit.
We thus obtain~\cite{kr1,krlead}
\beq
f_k(n,i)\approx\frac{(\ln z)^{k-1}}{z\,(k-1)!}.
\label{ufz}
\eeq

\subsection{Degree statistics of the whole network}

We now turn to the degree distribution of the whole network at time $n$,
$f_k(n)=\prob\{k(n)=k\}$,
where $k(n)$ stands for the degree of an unspecified node.
We have
\beq
f_k(n)=\frac{1}{n}\sum_{i=1}^n f_k(n,i).
\eeq
The corresponding generating polynomials,
\beq
F_n(x)=\bigmean{x^{k(n)}}=\sum_{k=1}^nf_k(n)x^k
=\frac{1}{n}\sum_{i=1}^n F_{n,i}(x),
\eeq
obey the recursion
\beq
(n+1)F_{n+1}(x)=(x+n-1)F_n(x)+x,
\label{fnrec}
\eeq
or equivalently
\beq
(n+1)f_k(n+1)=f_{k-1}(n)+(n-1)f_k(n)+\delta_{k,1},
\label{fknrec}
\eeq
with initial conditions given in Table~\ref{tabledef}, i.e.,
\beq
f_k^\A(1)=\delta_{k,0},\quad
f_k^\B(1)=\delta_{k,1}.
\label{f1init}
\eeq

The recursion~(\ref{fnrec}) has a non-polynomial solution, independent of $n$,
\beq
F_\st(x)=\frac{x}{2-x},
\eeq
describing the stationary
degree distribution on an infinitely large network:
\beq
f_{k,\st}=\frac{1}{2^k}\quad(k\ge1).
\label{fstat}
\eeq
The solution of~(\ref{fnrec}) reads
\beq
\matrix{
F_n^\A(x)
=\frad{x}{2-x}+\frad{2(1-x)}{2-x}\,\frad{\Gamma(x+n-1)}{n!\Gamma(x)},\cr\cr
F_n^\B(x)
=\frad{x}{2-x}+\frad{x(1-x)}{2-x}\,\frad{\Gamma(x+n-1)}{n!\Gamma(x)}.}
\label{fnres}
\eeq

The polynomials $F_n^\A(x)$ and $F_n^\B(x)$
have respective degrees $n-1$ and~$n$.
The first of them which are not listed in Table~\ref{tabledef} read
\beq
\matrix{
F_3^\A(x)=\frac{1}{3}\,x(x+2),\hfill&
F_3^\B(x)=\frac{1}{6}\,x(x^2+2x+3),\hfill\cr\cr
F_4^\A(x)=\frac{1}{12}\,x(x^2+4x+7),\hfill&
F_4^\B(x)=\frac{1}{24}\,x(x+3)(x^2+x+4).\hfill}
\eeq
The degree $k(n)$ at time $n$ ranges from the
minimal value 1 to the maximal value $n-1$ (Case~A) or $n$ (Case~B).
These extremal values occur with the following probabilities $(n\ge2)$
\beq
f_1^\A(n)=\frac{1}{2}+\frac{1}{n(n-1)},\quad
f_1^\B(n)=\frac{1}{2},\quad
f_{n-1}^\A(n)=\frac{2}{n!},\quad
f_n^\B(n)=\frac{1}{n!}.
\label{ffac}
\eeq

\begin{figure}
\begin{center}
\includegraphics[angle=90,width=.7\linewidth]{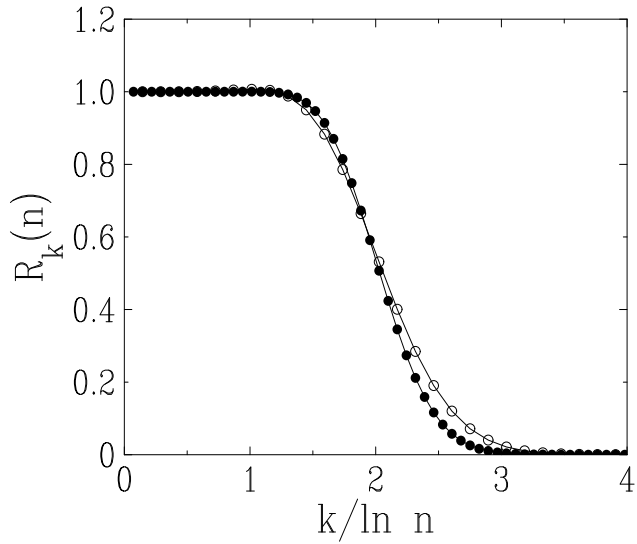}
\caption{\label{auscaling}
Plot of the ratios $R_k(n)$ against $k/\ln n$ (see~(\ref{rdef})),
for the UA model with initial condition~A,
at times $n=10^3$ (empty symbols) and $n=10^6$ (full symbols).}
\end{center}
\end{figure}

We now turn to the finite-size scaling behavior of the degree distribution
when both $k$ and $n$ are large.
As anticipated in the Introduction,
it is to be expected that the probabilities $f_k(n)$ are close to their limits
$(f_k(n)\approx f_{k,\st})$ for $n$ large at fixed degree $k$,
and more generally in the stationary regime
where~$k$ is much smaller than some characteristic
crossover degree $k_\star(n)$.
Conversely, the probabilities $f_k(n)$ are expected to be negligible
$(f_k(n)\ll f_{k,\st})$ for $k$ large enough at fixed time $n$,
and more generally in the large-deviation regime where $k_\star(n)\ll k\sim n$.
The crossover scale $k_\star(n)$ can be estimated as
$k_\star(n)\approx\mean{k_1(n)}$ (see~(\ref{kave})).
Nodes with highest degrees are indeed typically expected to be the oldest ones.
An alternative route consists in using the argument
of extreme value statistics alluded to in the Introduction:
the largest degree $k_\star$ at time $n$
is such that the stationary probability of having $k\ge k_\star$
is of order $1/n$.
Both approaches consistently yield
\beq
k_\star(n)\sim\ln n.
\eeq
Finite-size effects are best revealed by considering the ratios
\beq
R_k(n)=\frac{f_k(n)}{f_{k,\st}}=2^k f_k(n).
\label{rdef}
\eeq
These ratios are expected to fall off to zero
for $k$ of the order of $k_\star(n)\sim\ln n$.
Figure~\ref{auscaling} shows a plot of the ratios $R_k(n)$ against $k/\ln n$,
for times $n=10^3$ and $n=10^6$ in Case~A.
Numerically exact values of the $f_k(n)$ are obtained
by iterating~(\ref{fknrec}).
A steeper and steeper crossover is clearly observed.

In order to get some quantitative information on the observed crossover,
it is advantageous to introduce the differences
$d_k(n)=R_{k-1}(n)-R_k(n)$ for $k\ge2$,
completed by $d_1(n)=1-R_1(n)$, i.e., $R_0(n)=1$.
Although the $d_k(n)$ are not positive, most of them are,
and they sum up to unity, so that it is tempting to think of them
as a narrow probability distribution living in the crossover region.
The generating function of the $d_k(n)$ reads
\beq
D_n(x)=\sum_{k\ge1}d_k(n)x^k=(x-1)F_n(2x)+x.
\label{phires}
\eeq
The above picture suggests to define the crossover scale as the first moment
\beq
k_\star=\mu(n)=\sum_{k\ge1}kd_k(n)=D_n'(1),
\eeq
and the squared width of the crossover front as the variance
\beq
\sigma^2(n)=\sum_{k\ge1}k^2d_k(n)-\mu(n)^2=D_n''(1)+\mu(n)-\mu(n)^2.
\eeq
Equations~(\ref{fnres}),~(\ref{phires}) yield
\beq
\mu^\A(n)=2H_n\approx2(\ln n+\euler),\quad
\mu^\B(n)=2H_n+1\approx2(\ln n+\euler)+1,
\eeq
and
\beq
\sigma^2(n)=2H_n-4H^\de_n\approx2(\ln n+\euler-\pi^2/3),
\eeq
the latter result being independent of the initial condition.

The crossover scale is thus $k_\star\approx2\ln n$,
whereas the width of the crossover front grows as
$\sigma(n)\approx(2\ln n)^{1/2}$.
These predictions are in agreement with the observations which can be made
on Figure~\ref{auscaling},
namely that the crossover takes place around $k/\ln n=2$,
and that it becomes steeper at larger times,
as its relative width falls off, albeit very slowly, as $(\ln n)^{-1/2}$.

\begin{figure}
\begin{center}
\includegraphics[angle=90,width=.7\linewidth]{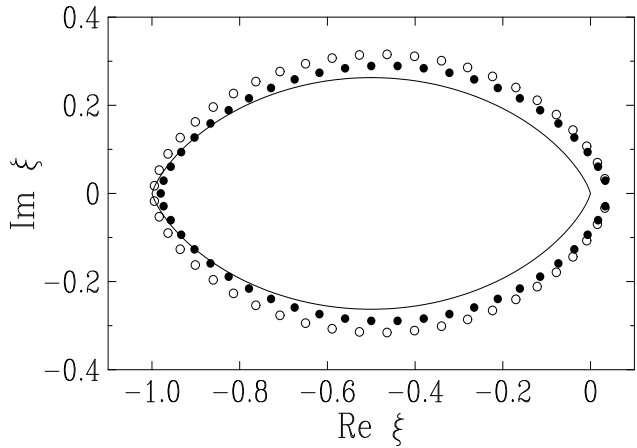}
\caption{\label{auzeros}
Plot of the non-trivial zeros of the polynomials $F_n(x)$ for the UA model,
in the complex plane of the rescaled variable $\xi=x/n$.
Symbols: zeros for $n=50$ in Case~A (empty symbols) and Case~B (full symbols).
Line: limiting curve with equation~(\ref{lens}).}
\end{center}
\end{figure}

Another illustration of finite-size effects is provided
by the complex zeros of the polynomials $F_n(x)$.
The location of these zeros indeed shows
how fast the degree distribution of finite networks,
encoded in the polynomials $F_n(x)$, converges to the stationary distribution,
encoded in the function $F_\st(x)$.
For $n\ge2$, $F_n^\A(x)$ and $F_n^\B(x)$ have one trivial zero at $x=0$,
and respectively $n-2$ and $n-1$ non-trivial ones.
The explicit expressions~(\ref{fnres}) allow one to find
the asymptotic locus of the zeros as follows.
The most rapidly varying part of these results is the rightmost ratio,
so that the zeros are asymptotically located on the curve with equation
$\abs{\Gamma(x+n-1)/(n!\Gamma(x))}=1$.
Setting
\beq
x=n\xi,
\label{xidef}
\eeq
and using Stirling's formula, we can recast the above estimate as
\beq
\Re{\left[(1+\xi)\ln(1+\xi)-\xi\ln\xi\right]}=0.
\label{lens}
\eeq
The non-trivial zeros of the polynomials $F_n(x)$
are thus predicted to escape to infinity linearly with time $n$.
Once rescaled by $n$ according to~(\ref{xidef}),
they accumulate onto a well-defined limiting curve in the complex $\xi$-plane.
This curve, with equation~(\ref{lens}),
has the shape of a lens connecting the points~$-1$ and $0$.
Figure~\ref{auzeros} illustrates this result with data at time $n=50$
for both initial conditions.
The polynomials $F_n(x)$ converge to the stationary function $F_\st(x)$
whenever the complex ratio $\xi=x/n$ lies within the lens.
Otherwise they diverge exponentially with $n$.

A related issue concerns the behavior of the probability $f_k(n)$
of having a very large degree, of order $k\sim n$,
much larger than $k_\star(n)\sim\ln n$.
Considering Case~A for definiteness,
the expression~(\ref{fnres}) leads to the exact contour-integral representation
\beq
f_k^\A(n)=\oint\frac{\d x}{2\pi\i\,x^{k+1}}
\left(\frac{x}{2-x}+\frac{2(1-x)}{2-x}\,\frac{\Gamma(x+n-1)}{n!\Gamma(x)}
\right).
\eeq
The presence of gamma functions suggests to look for a saddle point $x_\col$
proportional to $n$.
Setting $\zeta=k/n$, we indeed find $x_\col=n/v$, where $\zeta$ and
$v$ are related through
\beq
\zeta=\frac{\ln(v+1)}{v}.
\eeq
We thus obtain the following large-deviation estimate
\beq
f_k(n)\sim\exp\Bigl(-n\bigl(\zeta\ln n+S(\zeta)\bigr)\Bigr),
\label{uent}
\eeq
where the exponent has a usual contribution in $n$
and a less usual one in $n\ln n$.
The term linear in $n$ involves a large-deviation function $S(\zeta)$,
which is obtained in the following form, parametrized by $v$:
\beq
S(\zeta)=\frac{1}{v}\bigl(v\ln v-\ln v\ln(v+1)-(v+1)\ln(v+1)\bigr).
\eeq
This function decreases from $S(0)=0$ to $S(1)=-1$.
The resulting behavior at $\zeta=1$, i.e., $\exp(-n(\ln n-1))$,
is in agreement with the inverse factorial expressions~(\ref{ffac}).

\section{Linear preferential attachment: the Barab\'asi-Albert (BA) model}
\label{BA}

The Barab\'asi-Albert (BA) model is the simplest of the models
with preferential attachment:
each new node connects to earlier nodes with a probability proportional
to their degrees.
The probability that node $n$ connects to an earlier node $i$ thus reads
\beq
p_{n,i}=\frac{k_i(n-1)}{Z(n-1)},
\eeq
where $k_i(n-1)$ is the degree of node $i$ at time $n-1$,
i.e., before node $n$ enters the network.
The partition function in the denominator,
\beq
Z(n)=\sum_{i=1}^nk_i(n)=2L(n)
\eeq
(see~(\ref{sumr})), ensures that the attachment probabilities add up to unity.

In the following we analyze the BA model
along the lines of the previous section,
keeping consistent notations as much as possible.
The dependence of the attachment probability $p_{n,i}$
on the degree $k_i(n-1)$
however makes the problem more difficult than
the previous one of a uniform attachment.

\subsection{Degree statistics of a fixed node}

Let us again begin with the
distribution $f_k(n,i)=\prob\{k_i(n)=k\}$
of the degree of node $i$ at time $n$.

A first estimate of the degree $k_i(n)$ is provided by the following
recursion relation for the mean degree $\mean{k_i(n)}$,
which is a consequence of~(\ref{idef}):
\beq
\mean{k_i(n)}=\mean{k_i(n-1)}+\mean{p_{n,i}}
=\left(1+\frac{1}{Z(n-1)}\right)\mean{k_i(n-1)}.
\label{aveprod}
\eeq
In the scaling regime where both $i$ and $n$ are large,
using the expressions of the partition function given in Table~\ref{tabledef},
i.e.,
\beq
Z^\A(n)=2n-2,\quad Z^\B(n)=2n-1,
\label{zres}
\eeq
the above relation becomes the differential equation
\beq
\frac{\partial\mean{k_i(n)}}{\partial n}\approx\frac{\mean{k_i(n)}}{2n},
\eeq
which yields
\beq
\mean{k_i(n)}\approx\left(\frac{n}{i}\right)^{1/2}.
\label{bave}
\eeq

The generating polynomials $F_{n,i}(x)$ and $F_{n+1,i}(x)$
which encode the distribution of the degree of node $i$
at successive times $n$ and $n+1$ obey the recursion formula:
\beqa
F_{n+1,i}(x)
&=&\bigmean{x^{k_i(n+1)}}=\bigmean{x^{I_i(n+1)}x^{k_i(n)}}\nonumber\\
&=&\bigmean{(1+(x-1)p_{n+1,i})x^{k_i(n)}}\nonumber\\
&=&\bigmean{\left(1+\frac{x-1}{Z(n)}\,k_i(n)\right)x^{k_i(n)}},
\eeqa
i.e.,
\beq
F_{n+1,i}(x)=F_{n,i}(x)+\frac{x(x-1)}{Z(n)}\frac{\d F_{n,i}(x)}{\d x},
\label{fnirec}
\eeq
where $Z(n)$ is given by~(\ref{zres}).
The probabilities $f_k(n,i)$ themselves therefore obey the recursion
\beq
f_k(n+1,i)
=\frac{k-1}{Z(n)}f_{k-1}(n,i)+\left(1-\frac{k}{Z(n)}\right)f_k(n,i),
\eeq
with initial conditions~(\ref{f2init}).
The initial condition for Case~A should be taken at time $n=2$,
in order to avoid indeterminate expressions, as $Z^\A(1)=0$.

In order to solve the recursion~(\ref{fnirec}),
we perform the rational change of variable from $x$ to $u$ such that
\beq
u=\frac{x}{1-x},\quad x=\frac{u}{u+1},\quad
x(x-1)\frac{\d}{\d x}=-u\frac{\d}{\d u}.
\label{xtou}
\eeq
Introducing the notation $\w F_{n,i}(u)=F_{n,i}(x)$,
the recursion~(\ref{fnirec}) reads
\beq
\w F_{n+1,i}(u)=\w F_{n,i}(u)-\frac{u}{Z(n)}\frac{\d\w F_{n,i}(u)}{\d u}.
\label{wfnirec}
\eeq
It is then advantageous to introduce the Mellin transform $M_{n,i}(s)$
of $\w F_{n,i}(u)$, defined as
\beq
M_{n,i}(s)=\int_0^\infty\w F_{n,i}(u)\,u^{-s-1}\,\d u.
\eeq
The inverse transform reads
\beq
\w F_{n,i}(u)=\int_\C\frac{\d s}{2\pi\i}\,M_{n,i}(s)\,u^s,
\label{invmel}
\eeq
where C is a vertical contour in the complex $s$-plane
whose position will be defined in a while.
The virtue of the Mellin transformation
is that the recursion~(\ref{wfnirec}) simplifies to
\beq
M_{n+1,i}(s)=\left(1-\frac{s}{Z(n)}\right)M_{n,i}(s),
\label{mfnirec}
\eeq
with initial condition $M_{i,i}(s)=X_0(s)$ for $i\ge2$, with
\beq
X_0(s)=\int_0^\infty x(u)\,u^{-s-1}\,\d u=\int_0^1 x^{-s}(1-x)^{s-1}\,\d x
=\frac{\pi}{\sin\pi s}
\label{xdef}
\eeq
for $0<\Re s<1$.
Hereafter the contour C is assumed to be in that strip.
We thus get $(i\ge2)$
\beq
\matrix{
M_{n,i}^\A(s)=\frad{\Gamma\!\left(n-\frac{s}{2}-1\right)\Gamma(i-1)}
{\Gamma\!\left(i-\frac{s}{2}-1\right)\Gamma(n-1)}\,X_0(s),\hfill\cr\cr
M_{n,i}^\B(s)=\frad{\Gamma\!\left(n-\frac{s}{2}-\frac{1}{2}\right)
\,\Gamma\!\left(i-\frac{1}{2}\right)}
{\Gamma\!\left(i-\frac{s}{2}-\frac{1}{2}\right)
\,\Gamma\!\left(n-\frac{1}{2}\right)}\,X_0(s).\hfill
}
\label{bfnires}
\eeq
These product formulas in the Mellin variable $s$
are reminiscent of~(\ref{fnires}).

The mean and the variance of the degree of node $i$ at time~$n$
can be extracted from these results as follows.
The identity~(\ref{devt}) yields
\beq
\w F_{n,i}(u)=1-\frac{\mean{k_i(n)}}{u}
+\frac{\mean{k_i(n)^2}+\mean{k_i(n)}}{2u^2}+\cdots
\eeq
as $u\to+\infty$.
Furthermore the coefficients of $1/u$ and $1/u^2$
are respectively the residues of $M_{n,i}(s)$ at $s=-1$ and $s=-2$.
We thus obtain
\beq
\matrix{
\mean{k^\A_i(n)}=\frad{\Gamma\!\left(n-\frac{1}{2}\right)\Gamma(i-1)}
{\Gamma\!\left(i-\frac{1}{2}\right)\Gamma(n-1)},\quad
\mean{k^\B_i(n)}=\frad{\Gamma(n)\,\Gamma\!\left(i-\frac{1}{2}\right)}
{\Gamma(i)\,\Gamma\!\left(n-\frac{1}{2}\right)}
}
\eeq
and
\beq
\matrix{
\var{k^\A_i(n)}=2\,\frad{n-1}{i-1}-\mean{k^\A_i(n)}^2-\mean{k^\A_i(n)},\hfill
\cr\cr
\var{k^\B_i(n)}=2\,\frad{2n-1}{2i-1}-\mean{k^\B_i(n)}^2-\mean{k^\B_i(n)}.\hfill
}
\eeq

In the scaling regime where both times $i$ and $n$ are large and comparable,
introducing the time ratio $z=n/i$ (see~(\ref{zdef})),
the above results yield
\beq
\mean{k_i(n)}\approx z^{1/2},\quad\var{k_i(n)}\approx z^{1/2}(z^{1/2}-1),
\label{bkin}
\eeq
irrespective of the initial condition.
The mean degree is in agreement with the estimate~(\ref{bave}).
The entire degree distribution can actually be derived in the scaling regime.
Equation~(\ref{bfnires}) indeed yields
\beq
M_{n,i}(s)\approx z^{-s/2}\,\frac{\pi}{\sin\pi s}.
\eeq
We thus obtain
\beq
F_{n,i}(x)\approx\frac{x}{x+z^{1/2}(1-x)}
\eeq
and finally
\beq
f_k(n,i)\approx z^{-1/2}\bigl(1-z^{-1/2}\bigr)^{k-1}.
\label{bfz}
\eeq
The degree distribution is therefore found to be asymptotically geometric,
irrespective of the initial condition~\cite{kr1,krlead}.

\subsection{Degree statistics of the whole network}

We now turn to the degree distribution $f_k(n)=\prob\{k(n)=k\}$,
where $k(n)$ stands for the degree of an unspecified node.

The generating polynomials $F_n(x)$ obey the recursion
\beq
(n+1)F_{n+1}(x)=nF_n(x)+n\frac{x(x-1)}{Z(n)}\frac{\d F_n(x)}{\d x}+x,
\label{bfnrec}
\eeq
where $Z(n)$ is again given by~(\ref{zres}), and
with initial conditions given in Table~\ref{tabledef}.
The probabilities $f_k(n)$ themselves obey the recursion
\beq
(n+1)f_k(n+1)
=\frac{k-1}{Z(n)}nf_{k-1}(n)+\left(1-\frac{k}{Z(n)}\right)nf_k(n)+\delta_{k,1}.
\label{bfknrec}
\eeq

The first generating polynomials which depend on the attachment rule read
\beq
\matrix{
F_3^\A(x)=\frac{1}{3}\,x(x+2),\hfill&
F_3^\B(x)=\frac{1}{9}\,x(2x^2+2x+5),\hfill\cr\cr
F_4^\A(x)=\frac{1}{8}\,x(x^2+2x+5),\hfill&
F_4^\B(x)=\frac{1}{60}\,x(6x^3+8x^2+11x+35).\hfill}
\eeq

The stationary degree distribution $f_{k,\st}$
can be determined as the solution of~(\ref{bfknrec})
which becomes independent of $n$ for large $n$.
We thus get
\beq
(k+2)f_{k,\st}=(k-1)f_{k-1,\st}+2\delta_{k,1},
\eeq
hence~\cite{dms1,krl,kr1}
\beq
f_{k,\st}=\frac{4}{k(k+1)(k+2)}.
\label{bfstat}
\eeq
An alternative approach consists in looking for the
asymptotic generating function $F_\st(x)$
as the solution of~(\ref{bfnrec}) which becomes independent of $n$
for large~$n$.
We thus obtain the differential equation
\beq
x(1-x)F'_\st(x)+2F_\st(x)=2x,
\eeq
which has for solution
\beq
F_\st(x)=3-\frac{2}{x}-\frac{2(1-x)^2}{x^2}\ln(1-x).
\label{fst}
\eeq
Expanding this result as a power series in $x$
allows one to recover~(\ref{bfstat}).

The recursion~(\ref{bfnrec}) for the generating polynomials $F_n(x)$
can be solved along the lines of the above solution
of the recursion~(\ref{fnirec}).
The Mellin transforms $M_n(s)$ of the functions $\w F_n(u)=F_n(x)$
obey the recursion
\beq
(n+1)M_{n+1}(s)=\left(1-\frac{s}{Z(n)}\right)nM_n(s)+X_0(s),
\label{mfnrec}
\eeq
with initial condition $M_2^\A(s)=M_1^\B(s)=X_0(s)$.
Equation~(\ref{mfnrec}) has a special solution
\beq
M_n(s)=\frac{Z(n)X_0(s)}{(s+2)n},
\label{mspec}
\eeq
whereas the general solution of the homogeneous equation
shares the $n$-de\-pen\-den\-ce of the expressions~(\ref{bfnires}).
We thus obtain
\beq
\matrix{
M_n^\A(s)=\frad{2X_0(s)}{(s+2)n}
\left(n-1+(s+1)\frad{\Gamma\!\left(n-\frac{s}{2}-1\right)}
{\Gamma\!\left(1-\frac{s}{2}\right)\,\Gamma\!\left(n-1\right)}\right),\hfill\cr\cr
M_n^\B(s)=\frad{X_0(s)}{(s+2)n}\left(2n-1
+(s+1)\frad{\sqrt{\pi}\,\Gamma\!\left(n-\frac{s}{2}-\frac{1}{2}\right)}
{\Gamma\!\left(\frac{1}{2}-\frac{s}{2}\right)\,\Gamma\!\left(n-\frac{1}{2}\right)}
\right).\hfill
}
\label{bfnres}
\eeq
The common stationary limit of both expressions,
\beq
M_\st(s)=\frac{2X_0(s)}{s+2},
\eeq
is proportional to the special solution~(\ref{mspec}).
Recalling~(\ref{xdef}), the inverse Mellin transform of the above result,
\beq
\w F_\st(u)=1-\frac{2}{u}+\frac{2}{u^2}\ln(u+1),
\eeq
is equivalent to~(\ref{fst}).

The results~(\ref{bfnres}) allow one to investigate,
at least in principle, every feature of the degree distribution $f_k(n)$.
Let us take the example of the probability $f_1(n)$
for a node to have degree one.
The inverse formula~(\ref{invmel}) shows that this probability
is equal to minus the residue of $M_n(s)$ at $s=1$.
The nature of the subleading corrections to the stationary value $f_{1,\st}=2/3$
depends on the initial condition.
For Case~A we obtain ($n\ge2$)
\beq
f_1^\A(n)=\frad{2(n-1)}{3n}
+\frad{4\,\Gamma\!\left(n-\frac{3}{2}\right)}{3\sqrt{\pi}\,n\Gamma(n-1)}
=\frad{2}{3}-\frad{2}{3n}+\frad{4}{3\sqrt{\pi}\,n^{3/2}}+\cdots
\label{af1}
\eeq
More generally, all the probabilities $f_k(n)$ exhibit
a singular correction in $n^{-3/2}$.
Case~B has the remarkable property that all the probabilities $f_k(n)$
are rational functions of time $n$.
Their expansion at large $n$ therefore only involves integer powers of $1/n$.
We have e.g.
\beq
\matrix{
f_1^\B(n)=\frad{2n-1}{3n}=\frad{2}{3}-\frad{1}{3n},\hfill\cr\cr
f_2^\B(n)=\frad{n^2-2n+3}{3n(2n-3)}
=\frad{1}{6}-\frad{1}{12n}+\frad{3}{8n^2}+\cdots\hfill
}
\label{bf1}
\eeq

We now turn to the finite-size scaling behavior
of the degree distribution when both $k$ and $n$ are large.
The crossover scale $k_\star(n)$ can again be estimated
either using~(\ref{bave}) or by the argument of extreme value statistics.
Both approaches consistently yield
\beq
k_\star(n)\sim n^{1/2}.
\eeq
We will now show that the degree distribution
obeys the multiplicative finite-size scaling law
\beq
f_k(n)\approx f_{k,\st}\,\Phi(y),\quad y=\frac{k}{n^{1/2}},
\label{fss}
\eeq
where the scaling function $\Phi(y)$ is non-universal,
in the sense that it depends on the initial condition~\cite{kr2,ws}.
The proof of the scaling behavior~(\ref{fss})
and the determination of the scaling functions $\Phi^\A(y)$
and $\Phi^\B(y)$ go as follows.
Let us start with Case~A.
The second term of the expression~(\ref{bfnres}) for $M_n^\A(s)$ scales as
a power law for large $n$:
\beq
M_{n,\sca}^\A(s)\approx\frac{2(s+1)X_0(s)}
{(s+2)\,\Gamma\!\left(1-\frac{s}{2}\right)}\,n^{-s/2-1}.
\eeq
The inverse Mellin transform of the latter formula,
\beq
\w F_{n,\sca}^\A(u)\approx\frac{1}{n}\int_\C\frac{\d s}{2\pi\i}
\,\frac{2(s+1)X_0(s)}{(s+2)\,\Gamma\!\left(1-\frac{s}{2}\right)}
\left(u/n^{1/2}\right)^s,
\eeq
describes the scaling behavior of $\w F_n^\A(u)$
in the regime where $u$ and $n$ are simultaneously large,
with $u/n^{1/2}$ fixed.
Finally, by inserting the above scaling estimate
into the contour-integral representation
\beq
f_k^\A(n)=\oint\frac{\d x}{2\pi\i}\,\frac{F_n^\A(x)}{x^{k+1}}
=\oint\frac{\d u}{2\pi\i}\,\frac{\w F_n^\A(u)(u+1)^{k-1}}{u^{k+1}},
\label{bcontour}
\eeq
permuting the order of integrals, opening up the $u$-contour and using
\beq
\int_C\frac{\d u}{2\pi\i}\,\frac{(u+1)^{k-1}}{u^{k-s+1}}
=\frac{\Gamma(k)}{\Gamma(s)\Gamma(k-s+1)},
\eeq
we obtain after some algebra the scaling form~(\ref{fss}), with
\beq
\Phi^\A(y)=1+\frac{2}{\sqrt{\pi}}
\int_\C\frac{\d s}{2\pi\i}\,\frac{s+1}{s+2}
\,\Gamma\!\left(\frac{1-s}{2}\right)\left(\frac{y}{2}\right)^{s+2}.
\eeq
Case~B can be dealt with along the same lines.
We thus get the similar expression
\beq
\Phi^\B(y)=1+\int_\C\frac{\d s}{2\pi\i}\,\frac{s+1}{s+2}
\,\Gamma\!\left(1-\frac{s}{2}\right)\left(\frac{y}{2}\right)^{s+2}.
\eeq
The above expressions can be evaluated by closing the contour to the right
and summing the residues at the poles of the gamma functions.
We thus get
\beq
\matrix{
\Phi^\A(y)=1+\frad{8}{\sqrt{\pi}}
\ds\sum_{m\ge0}\frad{(-1)^m(m+1)}{(2m+3)m!}\left(\frac{y}{2}\right)^{2m+3},
\hfill\cr\cr
\Phi^\B(y)=1+
\ds\sum_{m\ge0}\frad{(-1)^m(m+1)(2m+3)}{(m+2)!}\left(\frac{y}{2}\right)^{2m+4},
\hfill
}
\eeq
i.e., finally
\beq
\matrix{
\Phi^\A(y)=\erfc\left(\frad{y}{2}\right)
+\frad{y}{\sqrt{\pi}}\left(1+\frad{y^2}{2}\right)\e^{-y^2/4},\hfill\cr\cr
\Phi^\B(y)=\left(1+\frad{y^2}{4}+\frad{y^4}{8}\right)\e^{-y^2/4},\hfill
}
\label{bfss}
\eeq
where erfc denotes the complementary error function.
The above expression for $\Phi^\A$ can be found in~\cite{kr2,ws},
whereas that for $\Phi^\B$ has been shown in~\cite{dms2} to hold for a slightly
different attachment rule and initial condition.

\begin{figure}
\begin{center}
\includegraphics[angle=90,width=.7\linewidth]{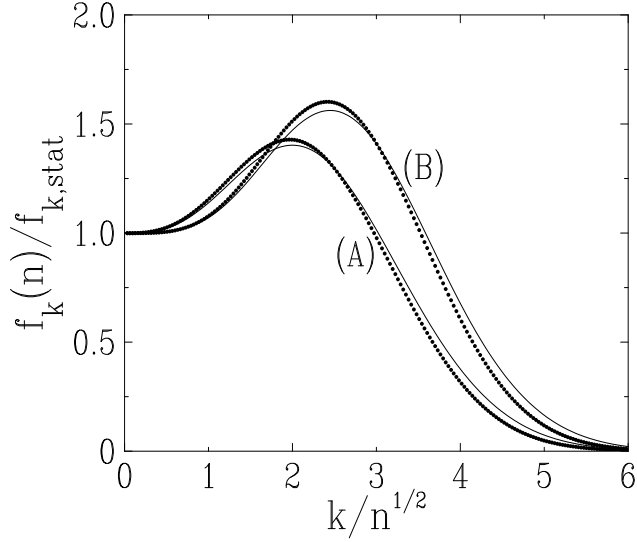}
\caption{\label{bscaling}
Plot of the ratios $f_k(n)/f_{k,\st}$
against the scaling variable $y=k/n^{1/2}$,
for the BA model at time $n=10^3$ (symbols) for both initial conditions.
Full lines: asymptotic scaling functions $\Phi^\A(y)$ and $\Phi^\B(y)$.}
\end{center}
\end{figure}

Figure~\ref{bscaling} shows a plot of the ratios $f_k(n)/f_{k,\st}$,
against the scaling variable $y=k/n^{1/2}$,
at time $n=10^3$ for both initial conditions.
Exact values for the $f_k(n)$ are obtained by iterating~(\ref{bfknrec}).
The data are well described by the predicted finite-size scaling functions
$\Phi^\A(y)$ and $\Phi^\B(y)$, shown as full lines.

Both scaling functions share similar qualitative features.
They start from the value 1 at $y=0$,
increase to a maximum, which is reached for $y=2$ in Case~A
and for $y=\sqrt{6}$ in Case~B, and fall off as $\exp(-y^2/4)$.
They however differ at the quantitative level,
both at small and large values of $y$:
\beq
\matrix{
\Phi^\A(y)=1+\frad{y^3}{3\sqrt{\pi}}+\cdots,\hfill&
\Phi^\B(y)=1+\frad{3y^4}{32}+\cdots,\hfill\cr\cr
\Phi^\A(y)\approx\frad{y^3}{2\sqrt{\pi}}\,\e^{-y^2/4},\hfill&
\Phi^\B(y)\approx\frad{y^4}{8}\,\e^{-y^2/4}.\hfill
}
\label{phiasy}
\eeq
Apart from the additive constant 1,
the scaling functions $\Phi^\A$ and $\Phi^\B$ are respectively
an odd and an even function of $y$.
This is the transcription in the finite-size scaling regime
of the phenomenon underlined when discussing~(\ref{af1}) and~(\ref{bf1}).
In particular, the first correction term at small $y$ is in $y^3$ for $\Phi^\A$,
and in $y^4$ for $\Phi^\B$.

\begin{figure}
\begin{center}
\includegraphics[angle=90,width=.6\linewidth]{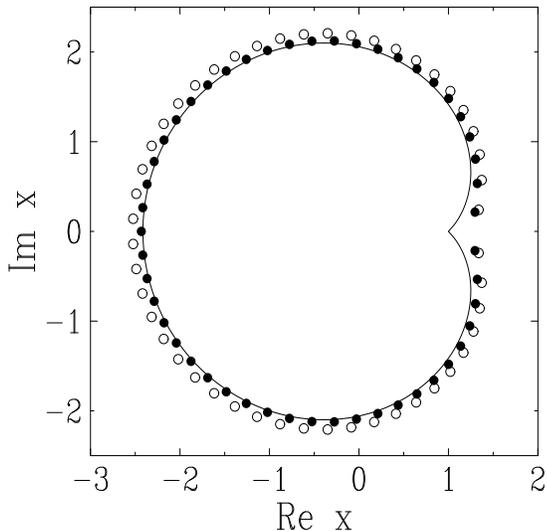}
\caption{\label{bzeros}
Plot of the non-trivial zeros of the polynomials $F_n(x)$
for the BA model in the complex $x$-plane.
Symbols: zeros for $n=50$ in Case~A (empty symbols) and Case~B (full symbols).
Line: limiting curve with equation~(\ref{bacurve}).}
\end{center}
\end{figure}

Let us again close up with the location
of the complex zeros of the polynomials $F_n(x)$.
Considering Case~A for definiteness,
the result~(\ref{bfnres}) can be recast as the exact formula
\beq
\w F_n^\A(u)-\frac{n-1}{n}\,\w F_\st(u)
=\frac{1}{n}\int_\C\frac{\d s}{\i}\,\frac{s+1}{s+2}
\,\frac{\Gamma\!\left(n-\frac{s}{2}-1\right)}
{\Gamma\!\left(1-\frac{s}{2}\right)\Gamma(n-1)}\,\frac{u^s}{\sin\pi s}.
\label{bfnc}
\eeq
The growth of this expression with $n$
for a fixed value of the complex variable $u$
can be investigated by means of the saddle-point approximation.
The presence of gamma functions again suggest to look for a saddle point
$s_\col$ proportional to~$n$.
Skipping details, let us mention that we find $s_\col\approx2n/(1-u^2)$,
so that the right-hand side of~(\ref{bfnc}) can be estimated as
\beq
\w F_{n,\sg}(u)\sim\left(1-\frac{1}{u^2}\right)^{-n},
\label{bexpo}
\eeq
with exponential accuracy.
The asymptotic locus of the complex zeros is then naturally given
by the condition that the above estimate neither falls off
nor grows exponentially.
We thus obtain $\abs{1-1/u^2}=1$.
The relevant part of this locus
can be parametrized by an angle $0\le\theta\le2\pi$~as
\beq
u=\left(1-\e^{-\i\theta}\right)^{-1/2},\quad
x=\frac{1}{1-\left(1-\e^{-\i\theta}\right)^{1/2}}.
\label{bacurve}
\eeq
This closed curve in the $x$-plane has a cusp at the point $x=1$,
corresponding to the scaling regime, with a right opening angle.
We have indeed $x-1\approx(\e^{\i\pi/2}\theta)^{1/2}$ as $\theta\to0$.
Figure~\ref{bzeros} illustrates this result with data at time $n=50$
for both initial conditions.
The polynomials $F_n(x)$ converge to the stationary series $F_\st(x)$
whenever the complex variable $x$ lies
within the closed curve shown on the figure.
Otherwise they diverge exponentially with $n$.

The exponential estimate~(\ref{bexpo}) has another virtue.
By inserting it
into the contour-integral representation~(\ref{bcontour}), we obtain
\beq
f_k(n)\sim\oint\frac{\d u}{2\pi\i}\left(\frac{u+1}{u}\right)^k
\left(1-\frac{1}{u^2}\right)^{-n}.
\label{fksaddle}
\eeq
This integral can in turn be investigated
by means of the saddle-point approximation.
The result is the following large-deviation estimate
\beq
f_k(n)\sim\exp(-n\,S(\zeta)),
\label{bent}
\eeq
where $\zeta=k/n$, and where the large-deviation function $S(\zeta)$ reads
\beq
S(\zeta)=(1-\zeta)\ln(1-\zeta)-(2-\zeta)\ln\frac{2-\zeta}{2}.
\eeq
The formula~(\ref{bent}) describes, with exponential accuracy,
the degree distribution in the whole large-deviation regime
where $k$ and $n$ are comparable.
The quadratic growth $S(\zeta)\approx\zeta^2/4$
at small $\zeta$ matches the fall-off of the finite-size scaling functions
$\Phi^\A(y)\sim\Phi^\B(y)\sim\exp(-y^2/4)$ (see~(\ref{phiasy})).
The maximal value $S(1)=\ln 2$ describes the fall-off $f_k(n)\sim2^{-n}$
of the probability of having
a degree $k$ equal to its maximal value ($k=n$ or $k=n-1$).

\section{The general preferential attachment (GPA) model}

We now consider the general preferential attachment (GPA) rule,
where the attachment probability to a node is proportional
to the sum $k_i(n)+c$ of the degree of the earlier node
and of an additive constant $c$,
representing the initial attractiveness of the node~\cite{dms1}.
This attachment rule interpolates between the uniform attachment rule,
which is recovered in the $c\to\infty$ limit,
and the BA model, which corresponds to $c=0$.
It can actually be continued on the other side of the BA model,
as $c$ can be chosen in the range $-1<c<\infty$.
The GPA model thus defined is scalefree for any finite value of $c$,
with the continuously varying exponents $\gamma=c+3$ and $\nu=1/(c+2)$.

The probability that node $n$ connects to an earlier node $i$ thus reads
\beq
p_{n,i}=\frac{k_i(n-1)+c}{Z(n-1)},
\label{gp}
\eeq
where $k_i(n-1)$ is the degree of node $i$ at time $n-1$,
and the partition function in the denominator,
\beq
Z(n)=\sum_{i=1}^n(k_i(n)+c)=2L(n)+cn
\label{zp}
\eeq
(see~(\ref{sumr})), ensures that the attachment probabilities add up to unity.

In the following we analyze the GPA model
along the very lines of the previous section.

\subsection{Degree statistics of a fixed node}

Let us again begin with the
distribution $f_k(n,i)=\prob\{k_i(n)=k\}$
of the degree of node $i$ at time $n$.

A first estimate of the degree $k_i(n)$ is provided by
the product formula~(\ref{aveprod}) for the mean degree $\mean{k_i(n)}$,
which still holds in the present case.
In the scaling regime where both $i$ and $n$ are large,
the latter relation becomes the differential equation
\beq
\frac{\partial\mean{k_i(n)}}{\partial n}\approx\frac{\mean{k_i(n)}+c}{(c+2)n},
\eeq
which yields
\beq
\mean{k_i(n)}\approx(c+1)\left(\frac{n}{i}\right)^{1/(c+2)}-c.
\label{gave}
\eeq
As anticipated, this expressions exhibits a power-law growth
with exponent $\nu=1/(c+2)$ in the range $0<\nu<1$.

The generating polynomials $F_{n,i}(x)$ and $F_{n+1,i}(x)$
associated with the degree of node $i$ at successive times
$n$ and $n+1$ obey the recursion formula:
\beqa
F_{n+1,i}(x)
&=&\bigmean{x^{k_i(n+1)}}=\bigmean{x^{I_i(n+1)}x^{k_i(n)}}\nonumber\\
&=&\bigmean{(1+(x-1)p_{n+1,i})x^{k_i(n)}}\nonumber\\
&=&\bigmean{\left(1+\frac{x-1}{Z(n)}\,(k_i(n)+c)\right)x^{k_i(n)}},
\eeqa
i.e.,
\beq
F_{n+1,i}(x)=F_{n,i}(x)+\frac{x-1}{Z(n)}
\left(c F_{n,i}(x)+x\frac{\d F_{n,i}(x)}{\d x}\right),
\label{gfnirec}
\eeq
where
\beq
Z^\A(n)=(c+2)n-2,\quad Z^\B(n)=(c+2)n-1.
\label{gzres}
\eeq
The probabilities $f_k(n,i)$ themselves therefore obey the recursion
\beq
f_k(n+1,i)
=\frac{k+c-1}{Z(n)}f_{k-1}(n,i)+\left(1-\frac{k+c}{Z(n)}\right)f_k(n,i),
\eeq
with initial conditions~(\ref{f2init}).

In order to solve the recursion~(\ref{gfnirec}),
we again perform the change of variable~(\ref{xtou}) from $x$ to $u$, and set
\beq
F_{n,i}(x)=(1-x)^{-c}\w F_{n,i}(u).
\eeq
The recursion~(\ref{gfnirec}) then reads
\beq
\w F_{n+1,i}(u)=\w F_{n,i}(u)-\frac{1}{Z(n)}
\left(c\w F_{n,i}(u)+u\frac{\d\w F_{n,i}(u)}{\d u}\right).
\label{gwfnirec}
\eeq
We then again introduce the Mellin transform $M_{n,i}(s)$ of $\w F_{n,i}(u)$,
so that the recursion~(\ref{gwfnirec}) simplifies to
\beq
M_{n+1,i}(s)=\left(1-\frac{s+c}{Z(n)}\right)M_{n,i}(s),
\label{gmfnirec}
\eeq
with initial condition $M_{i,i}(s)=X_c(s)$ for $i\ge2$, with
\beq
X_c(s)=\int_0^1 x^{-s}(1-x)^{s+c-1}\,\d x
=\frac{\Gamma(1-s)\Gamma(s+c)}{\Gamma(c+1)}
\label{gxdef}
\eeq
for $-c<\Re s<1$.
Hereafter the contour C is assumed to be in that strip.
We thus get $(i\ge2)$
\beq
\matrix{
M_{n,i}^\A(s)=\frad{\Gamma\!\left(n-\frac{s+c+2}{c+2}\right)
\,\Gamma\!\left(i-\frac{2}{c+2}\right)}
{\Gamma\!\left(i-\frac{s+c+2}{c+2}\right)\,\Gamma\!\left(n-\frac{2}{c+2}\right)}
\,X_c(s),\hfill\cr\cr
M_{n,i}^\B(s)=\frad{\Gamma\!\left(n-\frac{s+c+1}{c+2}\right)
\,\Gamma\!\left(i-\frac{1}{c+2}\right)}
{\Gamma\!\left(i-\frac{s+c+1}{c+2}\right)\,\Gamma\!\left(n-\frac{1}{c+2}\right)}
\,X_c(s),\hfill
}
\label{gfnires}
\eeq
These product formulas are a generalization of~(\ref{bfnires}).
The mean and the variance of the degree of node $i$ at time~$n$
can be extracted from these results as follows.
The identity~(\ref{devt}) now yields
\beq
\w F_{n,i}(u)=\frac{1}{u^c}-\frac{\mean{k_i(n)}+c}{u^{c+1}}
+\frac{\mean{k_i(n)^2}+(2c+1)\mean{k_i(n)}+c(c+1)}{2u^{c+2}}+\cdots
\eeq
as $u\to+\infty$.
Furthermore the coefficients of this expansion
are respectively the residues of $M_{n,i}(s)$
at $s=-c$, $s=-c-1$ and $s=-c-2$.
We thus obtain
\beq
\matrix{
\mean{k^\A_i(n)}=(c+1)\frad
{\Gamma\!\left(n-\frac{1}{c+2}\right)\,\Gamma\!\left(i-\frac{2}{c+2}\right)}
{\Gamma\!\left(i-\frac{1}{c+2}\right)\,\Gamma\!\left(n-\frac{2}{c+2}\right)}-c,
\hfill\cr\cr
\mean{k^\B_i(n)}=(c+1)\frad{\Gamma(n)\,\Gamma\!\left(i-\frac{1}{c+2}\right)}
{\Gamma(i)\,\Gamma\!\left(n-\frac{1}{c+2}\right)}-c\hfill
}
\eeq
and
\beq
\matrix{
\var{k^\A_i(n)}=(c+1)(c+2)\frad
{\Gamma(n)\,\Gamma\!\left(i-\frac{2}{c+2}\right)}
{\Gamma(i)\,\Gamma\!\left(n-\frac{2}{c+2}\right)}\hfill\cr\cr
{\hskip 47.5pt}-\mean{k^\A_i(n)}^2-(2c+1)\mean{k^\A_i(n)}-c(c+1),\hfill\cr\cr
\var{k^\B_i(n)}=(c+1)(c+2)\frad
{\Gamma\!\left(n+\frac{1}{c+2}\right)\,\Gamma\!\left(i-\frac{1}{c+2}\right)}
{\Gamma\!\left(i+\frac{1}{c+2}\right)\,\Gamma\!\left(n-\frac{1}{c+2}\right)}
\hfill\cr\cr
{\hskip 47.5pt}-\mean{k^\B_i(n)}^2-(2c+1)\mean{k^\B_i(n)}-c(c+1).\hfill
}
\eeq

In the scaling regime where both times $i$ and $n$ are large and comparable,
introducing the time ratio $z=n/i$ (see~(\ref{zdef})),
the above results yield
\beq
\mean{k_i(n)}\approx(c+1)z^{1/(c+2)}-c,\quad
\var{k_i(n)}\approx(c+1)z^{1/(c+2)}(z^{1/(c+2)}-1),
\label{gkin}
\eeq
irrespective of the initial condition.
The mean degree is in agreement with the estimate~(\ref{gave}).
The entire degree distribution can actually be derived in the scaling regime.
Equation~(\ref{gfnires}) indeed yields
\beq
M_{n,i}(s)\approx z^{-(s+c)/(c+2)}\,X_c(s).
\eeq
We thus obtain after some algebra
\beq
F_{n,i}(x)\approx\frac{x}{\left(x+z^{1/(c+2)}(1-x)\right)^{c+1}}
\eeq
and finally
\beq
f_k(n,i)\approx z^{-(c+1)/(c+2)}\bigl(1-z^{-1/(c+2)}\bigr)^{k-1}
\frac{\Gamma(k+c)}{\Gamma(k)\Gamma(c+1)}.
\label{gfz}
\eeq
This result allows one to recover both the Poissonian law~(\ref{ufz})
in the $c\to\infty$ limit and the geometric one~(\ref{bfz}) as $c=0$.

\subsection{Degree statistics of the whole network}

We now turn to the degree distribution of the whole network at time $n$,
$f_k(n)=\prob\{k(n)=k\}$,
where $k(n)$ stands for the degree of an unspecified node.

The generating polynomials $F_n(x)$ obey the recursion
\beq
(n+1)F_{n+1}(x)=nF_n(x)+\frac{n(x-1)}{Z(n)}
\left(c F_n(x)+x\frac{\d F_n(x)}{\d x}\right)+x,
\label{gfnrec}
\eeq
where $Z(n)$ is given by~(\ref{gzres}), and
with initial conditions given in Table~\ref{tabledef}.
The probabilities $f_k(n)$ themselves obey the recursion
\beq
(n+1)f_k(n+1)
=\frac{k+c-1}{Z(n)}nf_{k-1}(n)
+\left(1-\frac{k+c}{Z(n)}\right)nf_k(n)+\delta_{k,1}.
\label{gfknrec}
\eeq

The first generating polynomials which depend on the attachment rule read
\beq
\matrix{
F_3^\A(x)=\frac{1}{3}\,x(x+2),\hfill\cr\cr
F_3^\B(x)=\frac{1}{3(2c+3)}\,x\left((c+2)x^2+2(c+1)x+3c+5\right),\hfill\cr\cr
F_4^\A(x)=\frac{1}{4(3c+4)}\,x\left((c+2)x^2+4(c+1)x+7c+10\right),\hfill\cr\cr
F_4^\B(x)=\frac{1}{4(2c+3)(3c+5)}\,x
\bigl((c+2)(c+3)x^3+4(c+1)(c+2)x^2\hfill\cr
{\hskip 110pt}+(c+1)(7c+11)x+(3c+5)(4c+7)\bigr).\hfill
}
\eeq

The stationary degree distribution $f_{k,\st}$
can be determined as the solution of~(\ref{gfknrec})
which becomes independent of $n$ for large $n$.
We thus get
\beq
(k+2c+2)f_{k,\st}=(k+c-1)f_{k-1,\st}+(c+2)\delta_{k,1},
\label{fstrec}
\eeq
hence~\cite{dms1,krl}
\beq
f_{k,\st}=\frac{(c+2)\Gamma(2c+3)\Gamma(k+c)}{\Gamma(c+1)\Gamma(k+2c+3)}.
\label{gfstat}
\eeq
This result has a power-law decay at large $k$:
\beq
f_{k,\st}\approx\frac{(c+2)\Gamma(2c+3)}{\Gamma(c+1)}\,k^{-(c+3)}.
\label{gfkas}
\eeq
An alternative approach consists in looking for the
generating function $F_\st(x)$
as the stationary solution of~(\ref{gfnrec}).
We thus obtain the differential equation
\beq
x(1-x)F'_\st(x)+(2c+2-cx)F_\st(x)=(c+2)x,
\eeq
which is equivalent to~(\ref{fstrec}).
The solution
\beq
F_\st(x)=\frac{(c+2)(1-x)^{c+2}}{x^{2c+2}}
\int_0^x\frac{y^{2c+2}}{(1-y)^{c+3}}\,\d y
\eeq
can be recast in terms of a hypergeometric function,
which boils down to elementary functions whenever $2c$ is an integer.

Throughout the regime where the degree $k$ and the parameter $c$
are both large and comparable,
the expression~(\ref{gfstat}) assumes a stationary large-de\-vi\-a\-tion form,
\beq
f_{k,\st}\sim\exp(-c\,\phi(\kappa)),
\eeq
where $\kappa=k/c$, and with
\beq
\phi(\kappa)=(\kappa+2)\ln(\kappa+2)-(\kappa+1)\ln(\kappa+1)-2\ln 2.
\eeq
The linear behavior $\phi(\kappa)\approx\kappa\ln 2$ as $\kappa\to0$
matches the exponential decay~(\ref{fstat})
of the stationary distribution in the UA model,
formally corresponding to $c\to\infty$,
whereas the logarithmic growth $\phi(\kappa)\approx\ln\kappa+1-2\ln 2$
as $\kappa\to\infty$ matches the power-law decay~(\ref{gfkas}).

The moments of the stationary distribution,
\beq
m_p=\sum_{k\ge1}k^p\,f_{k,\st},
\eeq
can be derived from~(\ref{fstrec}), which yields the recursion
\beq
(c+2-p)m_p=c+2+pcm_{p-1}+\sum_{q=0}^{p-2}{{p}\choose{q}}(m_{q+1}+cm_q).
\eeq
We thus get
\beq
\matrix{
m_0=1,\quad m_1=2,\quad m_2=\frad{2(3c+2)}{c},\hfill\cr\cr
m_3=\frad{2(13c^2+17c+6)}{c(c-1)},\quad
m_4=\frad{2(3c+2)(25c^2+33c+14)}{c(c-1)(c-2)},\hfill
}
\eeq
and so on.
The power-law decay~(\ref{gfkas})
implies that the moment $m_p$ is convergent for $c>p-2$.

The recursion~(\ref{gfnrec}) for the generating polynomials $F_n(x)$
can again be exactly solved for a finite time $n$.
The Mellin transforms $M_n(s)$ of the functions
$\w F_n(u)=(1-x)^c F_n(x)$ obey
\beq
(n+1)M_{n+1}(s)=\left(1-\frac{s+c}{Z(n)}\right)nM_n(s)+X_c(s),
\label{gmfnrec}
\eeq
with initial condition $M_2^\A(s)=M_1^\B(s)=X_c(s)$.
Equation~(\ref{gmfnrec}) has a special solution
\beq
M_n(s)=\frac{Z(n)X_c(s)}{(s+2c+2)n},
\label{gmspec}
\eeq
whereas the general solution of the homogeneous equation
shares the $n$-de\-pen\-den\-ce of the expressions~(\ref{gfnires}).
We thus get
\beq
\matrix{
M_n^\A(s)=\frad{X_c(s)}{(s+2c+2)n}\hfill\cr
{\hskip 36pt}\times
\left((c+2)n-2+2(s+c+1)\frad{\Gamma\!\left(n-\frac{s+c+2}{c+2}\right)
\,\Gamma\!\left(\frac{2c+2}{c+2}\right)}
{\Gamma\!\left(1-\frac{s}{c+2}\right)\,\Gamma\!\left(n-\frac{2}{c+2}\right)}\right)
,\hfill\cr\cr
M_n^\B(s)=\frad{X_c(s)}{(s+2c+2)n}\hfill\cr
{\hskip 36pt}\times
\left((c+2)n-1+(s+c+1)\frad{\Gamma\!\left(n-\frac{s+c+1}{c+2}\right)
\,\Gamma\!\left(\frac{c+1}{c+2}\right)}
{\Gamma\!\left(\frac{1-s}{c+2}\right)\,\Gamma\!\left(n-\frac{1}{c+2}\right)}\right)
.\hfill
}
\label{gfnres}
\eeq

In order to illustrate these general results,
let us again consider the probability $f_1(n)$ for a node to have degree one.
This probability is minus the residue of $M_n(s)$ at $s=1$.
For Case~A we obtain ($n\ge2$)
\beqa
f_1^\A(n)&=&\frad{1}{(2c+3)n}
\left((c+2)n-2+2(c+2)\frad{\Gamma\!\left(n-\frac{c+3}{c+2}\right)
\,\Gamma\!\left(\frac{2c+2}{c+2}\right)}
{\Gamma\!\left(\frac{c+1}{c+2}\right)\,\Gamma\!\left(n-\frac{2}{c+2}\right)}\right)
\nonumber\\
&=&\frac{1}{2c+3}
\left(c+2-\frac{2}{n}+\frac{2(c+2)\,\Gamma\!\left(\frac{2c+2}{c+2}\right)}
{\Gamma\!\left(\frac{c+1}{c+2}\right)}
\,n^{-\frat{2c+3}{c+2}}+\cdots\right),
\eeqa
whereas for Case~B we obtain ($n\ge2$)
\beq
f_1^\B(n)=\frac{(c+2)n-1}{(2c+3)n}.
\eeq
This rational expression for $f_1^\B(n)$ is however an exception.
The probabilities $f_k(n)$ indeed generically have
a singular correction in $n^{-(2c+3)/(c+2)}$
for both initial conditions,
whereas only $f_1^\B(n)$ and $f_2^\B(n)$ are rational functions of time $n$.

We now turn to the finite-size scaling behavior
of the degree distribution when both $k$ and $n$ are large.
The crossover scale $k_\star(n)$ can again be estimated
either using~(\ref{gave}) or by the argument of extreme value statistics.
Both approaches consistently yield
\beq
k_\star(n)\sim n^{1/(c+2)}.
\eeq
The degree distribution
obeys a finite-size scaling law of the form
\beq
f_k(n)\approx f_{k,\st}\,\Phi(y),\quad y=\frac{k}{n^{1/(c+2)}},
\eeq
where the scaling function $\Phi(y)$ again depends
on the initial condition~\cite{kr2,ws}.
The determination of the scaling functions $\Phi^\A(y)$
and $\Phi^\B(y)$ closely follows the steps of Section 3.2.
We thus obtain
\beq
\matrix{
\Phi^\A(y)=1+\frad{2\,\Gamma\!\left(\frac{2c+2}{c+2}\right)}{(c+2)\Gamma(2c+3)}
\int_\C\frad{\d s}{2\pi\i}\,\frad{s+c+1}{s+2c+2}
\frad{\Gamma(1-s)}{\Gamma\!\left(1-\frac{s}{c+2}\right)}\,y^{s+2c+2},\hfill\cr\cr
\Phi^\B(y)=1+\frad{\Gamma\!\left(\frac{c+1}{c+2}\right)}{(c+2)\Gamma(2c+3)}
\int_\C\frad{\d s}{2\pi\i}\,\frad{s+c+1}{s+2c+2}
\frad{\Gamma(1-s)}{\Gamma\!\left(\frac{1-s}{c+2}\right)}\,y^{s+2c+2}.\hfill
}
\label{gfss}
\eeq
By closing the contours to the right,
we can derive the following convergent series:
\beq
\matrix{
\Phi^\A(y)=1+\frad{2\,\Gamma\!\left(\frac{2c+2}{c+2}\right)}{(c+2)\Gamma(2c+3)}
y^{2c+3}\ds\sum_{m\ge0}\frad{(m+c+2)(-y)^m}
{(m+2c+3)m!\,\Gamma\!\left(1-\frac{m+1}{c+2}\right)},\hfill\cr\cr
\Phi^\B(y)=1+\frad{\Gamma\!\left(\frac{c+1}{c+2}\right)}{(c+2)\Gamma(2c+3)}
y^{2c+3}\ds\sum_{m\ge1}\frad{(m+c+2)(-y)^m}
{(m+2c+3)m!\,\Gamma\!\left(-\frac{m}{c+2}\right)}.\hfill
}
\label{gfsss}
\eeq
The above expression for $\Phi^\A$ can be found in~\cite{ws},
albeit not in a fully explicit form.
It is also worth mentioning that the finite-size scaling function derived
in~\cite{cb} for asymmetric growing networks is different from the above one
for generic values of the exponent $\nu=1/(c+2)$,
although it coincides for $\nu=1/2$ with our result~(\ref{bfss}) for $\Phi^\A$.

The expressions~(\ref{gfsss}) suggest that the derivatives
${\Phi^\A}'(y)$ and ${\Phi^\B}'(y)$ are somewhat simpler
than the functions themselves.
The factor $(m+2c+3)$ is indeed chased away from the denominators
under differentiation.
The resulting series can be resummed by means of the identities
\beq
\matrix{
\ds\sum_{m\ge0}\frad{(-y)^m}{m!\,\Gamma\!\left(1-\frac{m+1}{c+2}\right)}
=(c+2)\int_\C\frad{\d z}{2\pi\i}\,\e^{-yz+z^{c+2}},\hfill\cr\cr
\ds\sum_{m\ge1}\frad{(-y)^m}{m!\,\Gamma\!\left(-\frac{m}{c+2}\right)}
=y\int_\C\frad{\d z}{2\pi\i}\,\e^{-yz+z^{c+2}},\hfill
}
\eeq
which are known e.g.~in the theory of L\'evy stable laws.
We are thus left with the following alternative contour-integral expressions
for the derivatives:
\beq
\matrix{
{\Phi^\A}'(y)=\frad{2\,\Gamma\!\left(\frac{2c+2}{c+2}\right)}{\Gamma(2c+3)}
y^{2c+2}\int_\C\frad{\d z}{2\pi\i}(c+2-yz)\,\e^{-yz+z^{c+2}},\hfill\cr\cr
{\Phi^\B}'(y)=\frad{\Gamma\!\left(\frac{c+1}{c+2}\right)}{(c+2)\Gamma(2c+3)}
y^{2c+3}\int_\C\frad{\d z}{2\pi\i}(c+3-yz)\,\e^{-yz+z^{c+2}}.\hfill
}
\label{phiint}
\eeq

Both scaling functions start increasing from the value 1
according to the power laws
\beq
\matrix{
\Phi^\A(y)=1+\frad{2\,\Gamma\!\left(\frac{2c+2}{c+2}\right)}
{\Gamma(2c+4)\,\Gamma\!\left(\frac{c+1}{c+2}\right)}\,y^{2c+3}+\cdots,\hfill\cr\cr
\Phi^\B(y)=1+\frad{(c+3)}{2(c+2)^3\Gamma(2c+3)}\,y^{2c+4}+\cdots,\hfill
}
\eeq
go through a maximum, and fall off superexponentially as
\beq
\Phi^\A(y)\approx
2(c+2)C\Gamma\!\left(\frat{2c+2}{c+2}\right)\Psi(y),\quad
\Phi^\B(y)\approx
C\Gamma\!\left(\frat{c+1}{c+2}\right)y\,\Psi(y),
\label{phist}
\eeq
with
\beq
\Psi(y)=y^{2c+3-\!\frat{c}{2(c+1)}}
\,\exp\left(-(c+1)\left(\frac{y}{c+2}\right)^{\frat{c+2}{c+1}}\right)
\label{psist}
\eeq
and
\beq
C=\left[(2\pi(c+1))^{1/2}(c+2)^{\frat{2c+3}{2(c+1)}}\Gamma(2c+3)\right]^{-1}.
\eeq

\begin{figure}
\begin{center}
\includegraphics[angle=90,width=.7\linewidth]{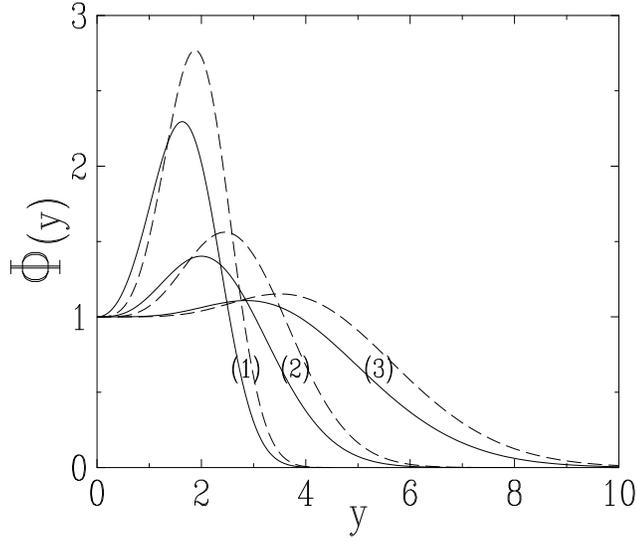}
\caption{\label{gscaling}
Plot of the scaling functions
$\Phi^\A(y)$ (full lines) and $\Phi^\B(y)$ (dashed lines) against $y$,
for (1) $c=-1/2$, i.e., $\nu=2/3$;
(2) $c=0$, i.e., $\nu=1/2$ (the BA model);
and (3) $c=1$, i.e., $\nu=1/3$.}
\end{center}
\end{figure}

Figure~\ref{gscaling} shows a plot of the scaling functions
$\Phi^\A(y)$ and $\Phi^\B(y)$
for (1) $c=-1/2$, i.e., $\nu=2/3$;
(2) $c=0$, i.e., $\nu=1/2$ (the BA model);
and (3) $c=1$, i.e., $\nu=1/3$.
The figure demonstrates that the scaling functions present
a high and narrow maximum for the smaller values of $c$,
and a direct crossover from 1 to 0 for the larger values of $c$.
These observations can be made quantitative by means of the pseudo-moments
\beq
\mu_p=-\int_0^\infty\Phi'(y) y^p\,\d y=p\int_0^\infty\Phi(y) y^{p-1}\,\d y.
\eeq
The integral formulas~(\ref{phiint}) allow one
to evaluate these quantities explicitly:
\beq
\matrix{
\mu_p^\A=\frad{(p+c+1)\,\Gamma\!\left(\frac{3c+4}{c+2}\right)\Gamma(p+2c+3)}
{(c+1)\,\Gamma\!\left(\frac{p+3c+4}{c+2}\right)\Gamma(2c+3)},\hfill\cr\cr
\mu_p^\B=\frad{\Gamma\!\left(\frac{c+1}{c+2}\right)\Gamma(p+2c+3)}
{\Gamma\!\left(\frac{p+c+1}{c+2}\right)\Gamma(2c+3)}.\hfill
}
\eeq

\noindent $\bullet$
For large values of $c$ (i.e., $c\to\infty$),
the model is close to the UA model.
The analysis of the scaling functions
will follow that of the ratios $R_k(n)$ in the UA model,
performed in Section~2.2.
The crossover value of $y$,
at which the functions exhibit a relatively sharp crossover from 1 to 0,
can be estimated as~$\mu_1$, i.e.,
\beq
\mu_1^\A=2c+2\euler+2+\cdots,\quad\mu_1^\B=2c+2\euler+3+\cdots,
\eeq
which grows as $2c$, irrespective of the initial condition.
Similarly, the squared width of the crossover region can be estimated as the
pseudo-variance $\sigma^2=\mu_2-\mu_1^2$, i.e.,
\beq
\sigma^{2\A}=2c+4\euler+2-2\pi^2/3+\cdots,\quad
\sigma^{2\B}=2c+4\euler+3-2\pi^2/3+\cdots,
\eeq
which also grows as $2c$, irrespective of the initial condition.

\noindent $\bullet$
For small values of $c$ (i.e., $c\to-1$),
the scaling functions exhibit a high and narrow peak around $y=1$.
The position of the peak can be estimated as $\mean{y}=\mu_2/(2\mu_1)$,
i.e., setting $c=-1+\eps$,
\beq
\mean{y}^\A=1+(3/2-\euler)\eps+\cdots,\quad
\mean{y}^\B=1+(2-\euler)\eps+\cdots,
\eeq
whereas the squared width of the peak can be estimated as
$\var{y}=(4\mu_1\mu_3-3\mu_2^2)/(12\mu_1^2)$, i.e.,
\beq
\matrix{
\var{y}^\A=\frad{5}{6}\,\eps+\frad{19-10\euler-\pi^2}{6}\,\eps^2+\cdots,\hfill\cr
\var{y}^\B=\frad{2}{3}\,\eps+\frad{22-8\euler-\pi^2}{6}\,\eps^2+\cdots,\hfill}
\eeq
and finally the area under the peak scales as $\mu_1$, i.e.,
\beq
\mu_1^\A=\frac{1}{\eps}+2-\euler+\cdots,\quad
\mu_1^\B=\frac{1}{\eps}+3-\euler+\cdots
\eeq
We are thus left with the picture of a narrow peak around $y=1$,
whose width shrinks as $\eps^{1/2}$ and whose height grows as $\eps^{-3/2}$.

\begin{figure}
\begin{center}
\includegraphics[angle=90,width=.6\linewidth]{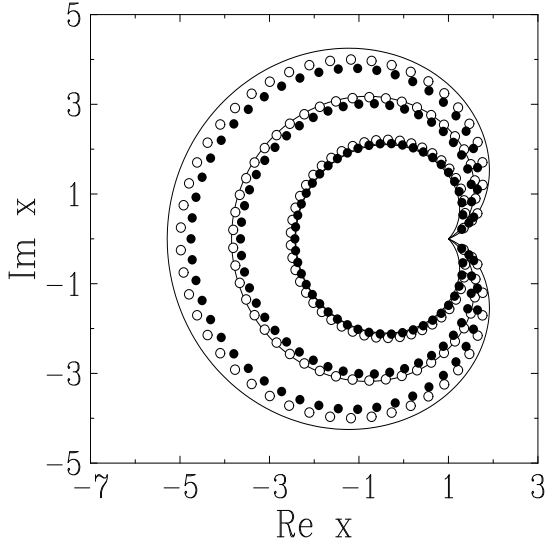}
\caption{\label{gzeros}
Plot of the non-trivial zeros of the polynomials $F_n(x)$
in the complex $x$-plane.
Symbols: zeros for $n=50$ in Case~A (empty symbols) and Case~B (full symbols).
Lines: limiting curves with equation~(\ref{gcurve}).
From the inside to the outside: $c=0$ (the BA model,
already shown in Figure~\ref{bzeros}), $c=1$ and $c=2$.}
\end{center}
\end{figure}

Let us close up this section with the location
of the complex zeros of the polynomials $F_n(x)$.
The derivation of the estimate~(\ref{bexpo}) can be generalized
to the present situation for arbitrary values of $c$.
We are thus left with
\beq
\w F_{n,\sg}(u)\sim\left(1-\frac{1}{(-u)^{c+2}}\right)^{-n},
\label{gexpo}
\eeq
again with exponential accuracy.
The asymptotic locus of the complex zeros is
therefore given by the condition $\abs{1-1/(-u)^{c+2}}=1$.
The relevant part of this locus
can be parametrized by an angle $0\le\theta\le2\pi$~as
\beq
u=\left(1-\e^{-\i\theta}\right)^{-1/(c+2)},\quad
x=\frac{1}{1-\left(1-\e^{-\i\theta}\right)^{1/(c+2)}}.
\label{gcurve}
\eeq
This closed curve in the $x$-plane has a cusp at the point $x=1$,
corresponding to the scaling regime, with an opening angle equal to $\pi/(c+2)$.
We have indeed $x-1\approx(\e^{\i\pi/2}\theta)^{1/(c+2)}$ as $\theta\to0$.
Figure~\ref{gzeros} illustrates this result with data at time $n=50$
for three values of $c$ and both initial conditions.

The exponential estimate~(\ref{gexpo})
can again be recast into a large-deviation estimate
for the probabilities $f_k(n)$ in the regime $k\sim n$, of the form
\beq
f_k(n)\sim\exp(-n\,S(\zeta)),
\label{gent}
\eeq
with $\zeta=k/n$.
The large-deviation function $S(\zeta)$ is obtained in parametric form:
\beq
\matrix{
\zeta=\frad{(c+2)(v-1)}{v^{c+2}-1},\hfill\cr\cr
S=\ln(v^{c+2}-1)
-\frad{c+2}{v^{c+2}-1}\left((v-1)\ln(v-1)+(v^{c+1}-1)v\ln v\right),
}
\eeq
where the parameter $v$ in the range $1<v<\infty$
is the opposite of the saddle-point value of $u$
in the contour integral generalizing~(\ref{fksaddle}).

The power-law behavior
\beq
S(\zeta)\approx
(c+1)\left(\frac{\zeta}{c+2}\right)^{\frat{c+2}{c+1}}
\eeq
at small $\zeta$ (corresponding to $v\to\infty$)
exactly matches the superexponential decay~(\ref{phist}),~(\ref{psist})
of the finite-size scaling functions $\Phi^\A(y)$ and $\Phi^\B(y)$.
The maximal value
\beq
S(1)=\ln(c+2),
\eeq
corresponding to $v\to1$, describes the exponential decay
$f_k(n)\sim(c+2)^{-n}$ of the probability of having
a degree $k$ equal to its maximal value ($k=n$ or $k=n-1$).

\section{Discussion}

In this paper we have presented a comprehensive study
of finite-size (i.e., finite-time) effects
on the degree statistics in growing networks.
We have considered models defined by stochastic attachment rules,
where nodes enter the network one at a time
and attach to one single earlier node,
so that the network has the topology of a tree.
The present study thus generalizes and extends many results
of References~\cite{dms1,dms2,krl,kr1,kr2,ws,mj}.

We have successively investigated the uniform attachment rule (UA),
the linear attachment rule of the Barab\'asi-Albert (BA) model,
and a general preferential attachment rule (GPA)
characterized by a continuous parameter $c>-1$,
representing the initial attractiveness of a node.
The UA and BA models are recovered as two special cases,
respectively corresponding to $c\to\infty$ and $c=0$.
The model is scalefree for any finite value of $c$,
with the continuously varying exponents $\gamma=c+3$ and $\nu=1/(c+2)$.
The continuous dependence of exponents on the parameter $c$,
and the dependence of finite-size scaling functions
on the initial condition (Case~A or Case~B in the present study),
are two illustrations of the lack of universality
which altogether characterizes the scaling behavior of growing networks.

The GPA rule is actually the most general one for which
the partition function $Z(n)$ (see~(\ref{zp})) is deterministic,
i.e., independent of the history of the network.
Whenever the attachment probability has a non-linear dependence on
the degree $k$,
the partition function becomes a history-dependent fluctuating quantity,
so that the analysis of size effects becomes far more difficult.
The general case of an arbitrary attachment rule,
growing either less or more rapidly than linearly with the degree,
has been considered in several works~\cite{krl,kr1,mj}.
Whenever the degree dependence of the attachment rule is asymptotically linear,
the resulting network is generically scalefree.
The determination of the degree exponent $\gamma$
is however a highly non-trivial
task in general (see~\cite{krl,kr1} for an explicit example).

The present study has underlined the key r\^ole
played by the typical value $k_\star(n)$ of the largest degree
in a finite network at time $n$.
In the UA model, $k_\star(n)$ grows logarithmically with time $n$.
The situation is more interesting in the scalefree case, i.e., for $c$ finite.
The largest degree $k_\star(n)$ grows as a subextensive power law
with exponent $\nu$,
and demarcates three regimes in the size-degree plane,
where finite-size (i.e., finite-time) effects
on the degree distribution $f_k(n)$ have different forms.

\noindent --
In the stationary regime ($k\ll k_\star(n)$),
the degree distribution is very close to the stationary one, $f_{k,\st}$.

\noindent --
In the finite-size scaling regime ($k\sim k_\star(n)$),
the degree distribution obeys
a multiplicative finite-size scaling law.
As already noticed in several earlier works,
the finite-size scaling function $\Phi$ depends
on the initial condition imposed on the network.
This lack of universality holds for all finite values of
the parameter~$c$.
Another feature of the finite-size scaling function is that
it increases from its initial value $\Phi(0)=1$, reaches a maximum,
and stays above unity for a range of values of its argument
$y=k/k_\star(n)$, before it eventually falls off to zero.
This non-monotonic overshooting behavior is however not mandatory.
In this respect it is worth recalling
the example of the so-called zeta urn model~\cite{dgc,zeta1,zetarev}.
This mean-field interacting particle system with multiple occupancies
possesses a continuous condensation transition at a finite critical density.
Its behavior right at the critical density shares
a high amount of similarity with the present problem,
including a power-law stationary distribution
with a continuously varying exponent, and finite-time scaling.
The same results have been shown
to apply to the dynamics of condensation
in the zero-range process (ZRP)~\cite{fsszrp}.
In the critical zeta urn and ZRP models, the finite-size scaling function
is a monotonically decreasing function, so that $\Phi(y)<1$ for all $y$.
This does not contradict the conservation of probability:
the excess probability is carried by smaller values of~$k$,
pertaining to the stationary regime.

\noindent --
In the large-deviation regime ($k_\star(n)\ll k\sim n$),
the degree distribution falls off exponentially in $n$.
At variance with the finite-size scaling law,
the corresponding large-deviation function
is independent of the initial condition.
The analysis of this regime has been shown to be closely related
to the locus of the complex zeros of the generating polynomials $F_n(x)$,
which have played a central r\^ole throughout this work.

To close up, it is to be hoped that some of the concepts and methods
used in the present work can be used to shed some new light
either to other observables in the network models considered here,
such as e.g.~the statistics of leaders and lead changes~\cite{krlead},
or to the degree statistics in more complex network models,
such as e.g.~the Bianconi-Barab\'asi (BB) model~\cite{bb1,bb2},
where attachment rules involve the competing effects
of dynamical variables (the node degrees)
and quenched disordered ones (the node fitnesses).
Depending on the a priori distribution of the random fitnesses,
the BB model may possess a low-temperature condensed phase.
Some features of the dynamics of the condensed phase have been
investigated recently,
both at zero temperature~\cite{usrecords},
where the model is intimately related to the statistics of records,
and at finite temperature~\cite{fb}.

\end{document}